\documentclass[journal,draftcls,onecolumn,twoside,11pt]{IEEEtran}
\normalsize

\usepackage{amsmath,amssymb,amsthm}
\usepackage{enumerate}
\usepackage{stfloats}
\usepackage{comment}
\usepackage{subfig}
\usepackage{lipsum}

\usepackage[dvipsnames]{xcolor}
\definecolor{myPink}{RGB}{255,105,183}

\usepackage[T1]{fontenc}
\usepackage{graphics} 
\usepackage{graphicx}
\usepackage{epsfig} 
\usepackage[mathscr]{euscript}
\usepackage{algorithm}
\usepackage[noend]{algpseudocode}
\makeatletter
\def\BState{\State\hskip-\ALG@thistlm}
\makeatother

\usepackage{tikz}
\usetikzlibrary{arrows,shapes,chains,matrix,positioning,scopes,patterns,calc,
decorations.markings,
decorations.pathmorphing,
}

\usepackage{pgfplots}
\pgfplotsset{compat=1.3}
\usepgflibrary{shapes}

\renewcommand{\epsilon}{\varepsilon}

\newcommand{\mb}[1]{\mathbf{#1}}

\newcommand{\mc}[1]{\mathcal{#1}}

\newcommand{\RNum}[1]{\uppercase\expandafter{\romannumeral #1\relax}}

\newcommand{\av}{\underline{{a}}}
\newcommand{\bv}{\underline{{b}}}

\newcommand{\vv}{\underline{{v}}}

\newcommand{\xv}{\underline{{x}}}

\newcommand{\yv}{\underline{{y}}}

\newcommand{\zv}{\underline{{z}}}

\def\Ktot{K_{\mathrm{tot}}}
\def\Ka{K_{\mathrm{a}}}

\def \Np{N_{\mathrm{p}}}
\def \Nc{N_{\mathrm{c}}}

\def \Bp{B_{\mathrm{p}}}
\def \Bc{B_{\mathrm{c}}}

\DeclareMathAlphabet{\mcl}{OMS}{cmsy}{m}{n}

\newlength\tikzwidth
\newlength\tikzheight

\newcommand{\coleq}{\mathrel{\mathop:}=}

\newif\ifjournal
\journalfalse

\def\Ktot{K_{\mathrm{tot}}}
\def\Ka{K_{\mathrm{a}}}
\def\Kb{K_{\mathrm{b}}}

\def\Pr{\mathrm{Pr}}

\usetikzlibrary{arrows,shapes,chains,matrix,positioning,scopes,patterns,calc,decorations.pathmorphing,shadows}
\usepackage[underline=true,rounded corners=false]{pgf-umlsd}

\IEEEoverridecommandlockouts\tikzstyle{cnode}=[circle,draw]
\tikzstyle{cgnode}=[circle,draw]
\tikzstyle{crnode}=[circle,draw]
\tikzstyle{conode}=[rectangle,draw]
\tikzstyle{cpnode}=[circle,draw]
\tikzstyle{rnode}=[rectangle,draw,outer sep=0pt]
\tikzstyle{fnode}=[rectangle,draw,rounded corners,fill=blue!25,align=center]
\tikzstyle{prnode}=[rectangle,rounded corners,fill=blue!50,text width=4.5em,text centered,outer sep=0pt]
\tikzstyle{prnodebig}=[rectangle,rounded corners,fill=blue!50,text width=7em,text centered,outer sep=0pt]
\tikzstyle{prnodesimple}=[rectangle,draw,text width=4.5em,text centered,outer sep=0pt]
\tikzstyle{bigsnake}=[snake=snake,segment amplitude=4mm, segment length=4mm, line after snake=5mm]
\tikzstyle{smallsnake}=[snake=snake,segment amplitude=0.7mm, segment length=4mm, line after snake=3mm]

\begin{document}
\title{Sparse IDMA: A Joint Graph-Based Coding Scheme for Unsourced Random Access}
\author{ Asit Kumar Pradhan, Vamsi Amalladinne, Avinash Vem,  Krishna R. Narayanan, and  Jean-Francois Chamberland \\
 Department of Electrical and Computer Engineering, Texas A\&M University
\thanks{
This material is based, partly, upon work supported by the National Science Foundation (NSF) under Grant No.~CCF-1619085. This work was presented in part at the IEEE Global Communications Conference, 2019.}
}
\maketitle	

\begin{abstract}
This article introduces a novel communication paradigm for the unsourced, uncoordinated Gaussian multiple access problem.
The major components of the envisioned framework are as follows.
The encoded bits of every message are partitioned into two groups. The first portion is transmitted using a compressive sensing scheme, whereas the second set of bits is conveyed using a multi-user coding scheme.
The compressive sensing portion is key in sidestepping some of the challenges posed by the unsourced aspect of the problem.
The information afforded by the compressive sensing is employed to create a sparse random multi-access graph conducive to joint decoding.
This construction leverages the lessons learned from traditional IDMA into creating low-complexity schemes for the unsourced setting and its inherent randomness.
Under joint message-passing decoding, the proposed scheme offers comparable performance  to existing low-complexity alternatives.
Findings are supported by numerical simulations.
\end{abstract}

\section{Introduction}
Recently, there has been a lot of interest in the design of novel access paradigms for uplink data transfers in IoT scenarios \cite{polyanskiy2017perspective,chen2017capacity,liva2011graph,paolini2015coded}.
These contributions propose a network with a very large number of devices, among which only a small subset, whose typical size is on the order of hundreds, are active at any given point in time. 
In \cite{polyanskiy2017perspective}, Polyanskiy poses the unsourced multiple access channel (MAC) problem where each active device wishes to transmit a $B$-bit message to a central base station and the base station is tasked with recovering the collection of $B$-bit messages communicated by the active users, without regard to the identity of the senders.
Therein, key finite block length (FBL) achievable bounds are derived for this setting. Since the publication of \cite{polyanskiy2017perspective}, there has been substantial interest in designing coding and decoding schemes with low complexity (polynomial in the number of message bits and the number of users) that perform close to the FBL bounds.
In \cite{ordentlich2017low}, Ordentlich and Polyanskiy report that traditional MAC coding schemes like ALOHA and treating interference as noise (TIN) exhibit performance far away from these FBL bounds.
They then introduce the first low-complexity algorithm tailored to the unsourced MAC setting.
In their scheme, the transmission period is divided into several slots, and each active user picks a random slot to transmit their message. Within each slot, a combination of compute-and-forward and forward error correction is employed for the $T$-user Gaussian MAC. 
While their scheme provides insights into the design of practical coding schemes for unsourced MAC, there remains a significant gap between the performance of this scheme and the FBL benchmarks.
Subsequently, several practical coding schemes have been proposed for the unsourced and uncoordinated MAC \cite{vem2017user,amalladinne2018coupled,Giuseppe,calderbank2018chirrup}.
Other related contributions, such as \cite{goseling2018sign}, present coding schemes for the uncoordinated random access channel which is closely related to the unsourced MAC.

In \cite{vem2017user}, Vem et al.\ devise a concatenated coding scheme with a slotted framework similar to \cite{ordentlich2017low} for the same problem.
This latter approach is, in essence, a per-user repetition scheme whereby codewords are sent over several slots.
The message corresponding to each active device is divided into two parts.
The first part is used to pick an interleaver for a low-density parity-check (LDPC) code, which is employed to encode the second part of the message.
The first part of the message is conveyed to the base station using compressive sensing (CS).
The second part is decoded using a per-slot message-passing decoder designed to recover data in the presence of up to $T-1$ users ($T \in\{2,4\}$).
The repetition pattern is a deterministic function of the user's message sequence.
A peeling decoder, which employs successive interference cancellation (SIC), works across slots to cancel the interference of successfully decoded messages.
This scheme is shown to perform significantly better than the scheme in \cite{ordentlich2017low} and is only around $6$~dB away from the FBL bounds.
In \cite{marshakov2019polar}, Marshakov et al.\  propose  using polar codes instead of LDPC codes to encode the second part of the message, which is decoded using a joint polar decoder. This further improves the error performance of the scheme in \cite{vem2017user}.

While the scheme in \cite{marshakov2019polar} uses time-division to sparsify active user's collisions, in \cite{pradhan2019polar}, Pradhan et al.\ use random spreading to alleviate multi-user interference.
The payload corresponding to each active user is split into two parts.
The first part acts as a preamble to choose a signature sequence from a codebook of sequences with good correlation properties.
The second part is encoded using a polar code whose frozen bits are dictated by the preamble.
The polar codeword is spread using the signature sequence picked by the first part of the message.
The decoder first employs an energy detector that uses correlation properties of the signature sequences to detect the list of spreading sequences used.
The preambles corresponding to active users are implicitly decoded in this step.
This information is passed to an minimum mean square error (MMSE) estimator that produces log-likelihood ratios (LLRs) for the latter part of messages based on the spreading sequences.
Finally, these LLRs are input to a single user polar list decoder that attempts to decode the latter part of messages treating interference as noise. The performance of the scheme is state-of-the-art when $\Ka < 175$; however, it does not scale well  with the number of active users. Another drawback of this scheme is that the decoding complexity is $\mathcal{O}(\Ka^3)$.

In \cite{amalladinne2018coupled}, Amalladinne et al.\ cast the unsourced MAC as a large compressive sensing problem.
They then construct a divide-and-conquer approach to obtain a pragmatic, low-complexity solution.
In their scheme, each active user's message is divided into several fragments, and these fragments are enhanced with redundancy. The coded sub-blocks are then encoded using a sensing matrix designed to recover the sub-blocks in the presence of noise using the non-negative least squares decoding algorithm. The recovered sub-blocks are then stitched together using the redundancy introduced during the encoding process. 
In \cite{Giuseppe}, Fengler et al.\  propose using approximate message passing (AMP) algorithm as the inner code in combination with the outer tree code in \cite{amalladinne2018coupled}. This scheme performs better than alternate schemes that preceded it. 
In \cite{CCSAMP}, Amalladinne et al.\ further improved the error performance by passing information between AMP and the outer code.
In addition, the work in \cite{calderbank2018chirrup} provides a very low complexity solution based on a chirp reconstruction algorithm. This complexity reduction, however, comes at the expense of error performance.


\subsection{Motivation and Contributions}
In this paper, we describe a novel low-complexity solution inspired from the scheme in~\cite{vem2017user} based on SIC. 
The proposed  coding scheme is based on decoding a joint Tanner graph for the unsourced MAC setting. Each active user's message is divided into two parts. The first part is used to schedule the transmission policy and pick a repetition factor for the latter part of the message. Similar to the scheme in \cite{vem2017user}, this part of the message is conveyed to the decoder using a CS framework. The other part of the message is encoded using an LDPC code.
Each bit of the LDPC codeword is then repeated a certain number of times, which is determined by the first part of the message.

We list below key features that distinguish our framework from prior art (details can be found in Section~\ref{Sec:SystemModel}).
\begin{enumerate}[i.]
\item In \cite{vem2017user}, messages are decoded on a per-slot basis, and copies are then peeled from other slots in the spirit of successive interference cancellation. In contrast, the approach we develop herein avoids the strategy of slotting-and-peeling altogether. A key contribution of this paper is to show that, when carefully designed, a single sparse joint Tanner graph that spans across all transmissions can provide substantial improvement in performance over the schemes in \cite{vem2017user,amalladinne2018coupled,Giuseppe}.

\item The scheme in \cite{vem2017user} relies on the existence of codes that achieve FBL capacity at the slot level. As the number of active users increases, the scheme in \cite{vem2017user} warrants that the slot length decrease. Designing FBL capacity achieving multi-user LDPC codes for such short block lengths becomes very challenging. 

\item Our proposed scheme can be interpreted as a sparse version of  IDMA \cite{ping2006interleave} adapted to the uncoordinated and unsourced MAC by using an additional compressed sensing part.  Unlike traditional IDMA, we carefully control the multi-user interference by keeping the transmissions sparse. Such sparsity is important in ensuring two key advantages: (a) the computational complexity of optimal soft-input soft-output demodulation is kept low, (b) the message passing decoding can perform efficiently for the large number of users and small message block lengths that are of interest in IoT. We derive the corresponding density evolution equations and optimize protograph based LDPC codes.
In the simulation results section, we show that the proposed approach significantly outperforms traditional IDMA for a large number of users.

\item The performance of the proposed scheme is comparable to that of the scheme in \cite{CCSAMP} although the two schemes are entirely different. In some parameter regimes, the proposed scheme is less complex than the scheme in \cite{CCSAMP}. Our scheme is also substantially less complex than the polar coding based scheme in \cite{pradhan2019polar} while the performance of the polar coding scheme is better.
\end{enumerate}

Throughout, we employ the following notation: $[a:b]$ denotes the set of integers from $a$ to $b$, including end points;
vectors are denoted by underlined symbols.

\section{System Model}
\label{Sec:SystemModel}
Let $\Ktot$ and $\Ka$ be the total number of users in the network and the number of active users, respectively.
Every active user has $B$ bits of information (or, equivalently, one of $M = 2^B$ indices) to be encoded and transmitted within a block of $N_{\mathrm{t}}$ uses of the channel. Let $W_i \in [0:M-1]$ be a random variable that represents the message index of the $i$th user and let $w_i$ be a realization of this random variable.
We assume that $W_i$ is uniformly distributed over $[0:M-1]$ and messages are independent from one another.

The observed signal vector at the receiver corresponding to the $N_{\mathrm{t}}$ channel uses can be written as
\begin{equation}
\yv = \sum_{i=1}^{\Ktot} s_i h_i\xv_{i}(w_i) + \zv,
\end{equation}
where $\xv_i(w_i) \in \mathbb{C}^{N_{\mathrm{t}}}$ is the signal transmitted by user~$i$, the additive noise is characterized by $\zv \sim \mc{CN}(0,\mb{I}_{N_{\mathrm{t}}})$, and $h_i \in \mathbb{C}$ are the fading coefficients which are independent of $\xv_i$ and $\zv$.
The Boolean indicator $s_i$ is defined as $s_i = 1$ if user~$i$ is active and $s_i = 0$ otherwise. We impose an average power constraint on the transmitted vectors when taken over all possible message indices, i.e., $\frac{1}{M}\sum_{w} \| \xv(w) \|^2\leq N_{\mathrm{t}}P$.
The energy-per-bit of the system is defined as $\frac{E_b}{N_0} \coleq \frac{N_{\mathrm{t}}P}{2B}$. The receiver produces an estimate $\mc{L}(\yv)$ of the list of messages.
As in \cite{polyanskiy2017perspective}, the probability of error is defined by
\begin{equation}\label{eqn:proboferrordefinition}
  P_e = \max_{|(s_1,\ldots,s_{\Ktot})| = \Ka} \frac{1}{\Ka} \sum_{i=1}^{\Ktot}  s_i\Pr\left( w_i \notin \mc{L}(\yv)|s_i=1 \right)
\end{equation}
where $|\cdot|$ denotes the Hamming weight.
The objective is to design a low-complexity encoding and decoding scheme with the least possible  $\frac{E_b}{N_0}$ such that $P_e \leq \epsilon$, where $\epsilon$ is the target error probability.

\section{Description of Proposed Scheme}
\label{Sec:ProposedScheme}
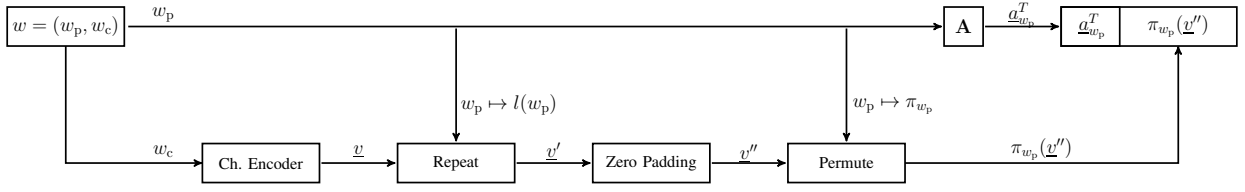
\begin{figure*}[t!]
  \centerline{\resizebox{\textwidth}{!}{\usetikzlibrary{arrows}
\begin{tikzpicture}
[draw=black, line width=1.25pt, >=stealth']
\def\fsize{\normalsize}
\def\fsizes{\scriptsize}
\def\ext{1}

\draw  (1.5,2) rectangle (4.5,1);
\node at (3,1.5) {\large Ch. Encoder};
\node (v4) at (4.4,1.5) {};
\node (v5) at (6.6,1.5) {};
\draw [->]  (v4) edge (v5);
\draw  (6.5,2) rectangle (9.5,1);
\node at (8,1.5) {\large Repeat};
\node (v6) at (9.4,1.5) {};
\node (v7) at (11.6,1.5) {};
\draw [->]  (v6) edge (v7);
\draw  (11.5,2) rectangle (14.5,1);
\node at (13,1.5) {\large Zero Padding};
\node (v8) at (14.4,1.5) {};
\node (v9) at (16.6,1.5) {};
\draw [->]  (v8) edge (v9);
\draw  (16.5,2) rectangle (19.5,1);
\node at (18,1.5) {\large Permute};
\draw  (-3.5,5.5) rectangle (-0.5,4.5);

\node (v1) at (-2,4.5) {};
\node (v2) at (-2,1.5) {};
\node (v3) at (1.5,1.5) {};

\draw  [->] plot[smooth, tension=0] coordinates {(v1) (v2) (v3)};
\node at (-2,5) {\Large $w = (w_{\mathrm{p}},w_{\mathrm{c}})$};

\draw  (20.5,5.5) rectangle (21.5,4.5);
\node at (21,5) {\Large  $\mathbf{A}$};
\node (v12) at (-0.55,5) {};
\node (v13) at (20.6,5) {};
\draw [->]  (v12) edge (v13);
\draw  (23.5,5.5) node (v10) {} rectangle (25,4.5);
\draw  (v10) rectangle (28,4.5);
\node (v14) at (23.6,5) {};
\node (v15) at (21.4,5) {};
\node (v16) at (26.5,4.5) {};
\node (v17) at (26.5,1.5) {};
\node (v18) at (19.5,1.5) {};
\draw [->]  (v15) edge (v14);
\draw  [->] plot[smooth, tension=0] coordinates {(v18) (v17) (v16)};
\node (v19) at (8,1.9) {};
\node (v20) at (8,5.1) {};
\node (v21) at (18,1.9) {};
\node (v22) at (18,5.1) {};
\draw [->]  (v20) edge (v19);
\draw [->]  (v22) edge (v21);
\node at (0.5,5.3) {\Large $w_{\mathrm{p}}$};
\node at (0.5,1.8) {\Large $w_{\mathrm{c}}$};
\node at (9.35,3) {\Large $w_\mathrm{p} \mapsto l({w_\mathrm{p}})$};
\node at (19.2,2.95) {\Large $w_\mathrm{p} \mapsto \pi_{w_\mathrm{p}}$};
\node at (24.3,5) {\Large $\underline{{a}}^T_{w_\mathrm{p}}$};
\node at (5.5,1.8) {\Large $\underline{v}$};
\node at (10.5,1.8) {\Large $\underline{v}'$};
\node at (15.5,1.75) {\Large $\underline{v}''$};
\node at (23,1.85) {\Large $\pi_{{w_\mathrm{p}}}(\underline{v}'')$};
\node at (26.5,5) {\Large $\pi_{{w_\mathrm{p}}}(\underline{v}'')$};
\node at (22.5,5.3) {\Large $\underline{{a}}^T_{w_\mathrm{p}}$};
\end{tikzpicture} }}
  \caption{This figure illustrates encoding process of the proposed scheme.}
 \label{fig:encodingscheme}
\end{figure*}

The overall schematic of the proposed scheme is illustrated in Fig.~\ref{fig:encodingscheme}. The parameters of the encoding process in our unsourced setting are independent of the user identity. So, our description of the encoding process is solely based on the message index; the encoding process is identical for every active user. 

\subsection{Encoder}
The encoder contains two components: a sensing matrix for a $\Ka$-sparse robust compressed sensing (CS) problem, and a multi-user channel code for the binary-input real-adder multiple-access channel.
The $N_{\mathrm{t}}$ channel uses available for communication are split between these two components: $\Np$ channel uses for the compressed sensing part ($\mathrm{p}$ denotes preamble) and $\Nc\coleq N_{\mathrm{t}}-\Np$ channel uses for the channel coding part. 
The $B$ bits to be transmitted are also split into two groups of $B_{\mathrm{p}}$ and $B_{\mathrm{c}} \coleq B-B_{\mathrm{p}}$ bits, respectively ($B_{\mathrm{p}} \ll B_{\mathrm{c}}$).
For convenience, we define $M_{\mathrm{p}} \coleq 2^{B_{\mathrm{p}}}$ and $M_{\mathrm{c}} \coleq 2^{B_{\mathrm{c}}}$.
Also, we denote the preamble and channel coding parts of the message index by $w_{\mathrm{p}}$ and $w_{\mathrm{c}}$.

For the CS portion of the encoding process, we consider a sensing matrix of the form $\mathbf{A} = \big[ \av_1~\av_2~\cdots~\av_{M_{\mathrm{p}}} \big] \in \mathbb{R}^{\Np\times M_{\mathrm{p}}}$ normalized to meet the power constraint, i.e. $\| \av_j \|_2^2 \le \Np P_1~\forall~1 \le j \le M_{\mathrm{p}}$.
An active user encodes its preamble message $w_{\mathrm{p}}$ into the column $\av_{w_{\mathrm{p}}}$ of $\mathbf{A}$.

The channel coding part of the message index $w_{\mathrm{c}}$ is first encoded into an $N$-bit codeword $\vv$ of an $(N,B_c)$ LDPC code $\mathcal{C}_{\text{LDPC}}$ and modulated using binary phase shift keying (BPSK).
The active user subsequently employs the many-to-one function $l:[1:M_\mathrm{p}] \rightarrow \{1, 2, \ldots, L\}$ to generate an integer $l({w_\mathrm{p}})$ based on $w_p$, and the LDPC codeword is repeated $l({w_\mathrm{p}})$ times.
The vector thus constructed takes the form $$\vv'=  \underbrace{\big[ \vv,\vv,\ldots,\vv \big]}_{l({w_{\mathrm{p}}})~ \text{copies}} .$$
Vector $\vv'$ is then padded with $N_c-Nl({w_{\mathrm{p}}})$ zeros to generate the $N_c$-length vector $\vv''=[\vv', 0, \ldots, 0]$
and normalized to satisfy the power constraint $\| \vv'' \|_2^2 \le N_tP-N_pP_1$.
At this stage, the preamble message $w_{\mathrm{p}}$ is again used to pick an interleaver $\pi_{{w_\mathrm{p}}}$ for the zero padded codeword $\vv''$.

Let ${\tilde{\vv}}_w$ be the codeword corresponding to message index $w=(w_{\mathrm{p}}, w_{\mathrm{c}})$. Then, ${\tilde{\vv}}_w$ is obtained by first permuting the zero-padded codeword $\vv''$ employing the permutation $\pi_{{w_\mathrm{p}}}$ and then inserting the $w_\mathrm{p}$th column of the sensing matrix $\mathbf{A}$ at the beginning of the permuted codeword, i.e.,
\begin{equation}\label{eqn:codeconstruction1}
  {{\tilde{\vv}}_w = [\av^T_{w_\mathrm{p}},\pi_{{w_\mathrm{p}}}(\vv'')]}.
\end{equation}
%
The key idea of the proposed construction is that zero-padding followed by interleaving the codeword $\vv'$ `sparsifies' the transmissions and reduces the interference in each use of the channel significantly, especially when $N \ll \Nc$.
Specifically, the average channel as seen at each time index is (approximately) a $\frac{1}{\Nc} \sum_{l=1}^{L} \nu_l Nl$-user Gaussian MAC rather than a $\Ka$-user Gaussian MAC, where $\nu_l$ denotes the fraction of users that employ repetition factor $l$.
This approach results in a superior performance, and it enables us to design a computationally efficient decoding algorithm.

\subsection{Decoder}
The overall decoder has two components.
The compressed sensing decoder recovers the preamble fragments, and concomitantly acquires the set of interleavers and repetition patterns picked by the active users.
A low-complexity message passing decoder then recovers the codewords sent over the $\Ka$-user Gaussian multiple access channel.

\subsubsection{Compressed Sensing Decoder}
\label{sec:CS_decoder}
The first $\Np$ received symbols can be written in vector form as
\begin{equation}
\label{eq:cross_correlation}
\yv_{\mathrm{p}} \coleq \yv[1:\Np] = \mathbf{A}\bv+\zv[1:\Np]
\end{equation}
where $\bv \in \mathbb{C}^{M_\mathrm{p}}$ is a $\Ka$-sparse vector whose support indicates the set of transmitted preamble messages and entries indicate the fading coefficients of the corresponding users. 
We first run a generic CS decoder , which yields estimate $\hat{\bv}$ of $\bv$.
Yet, we emphasize that this does not guarantee an output signal of the required sparsity  (as we know a priori from the structure of the problem). To address this issue, we sort the candidates and choose the list of the top $\Kb$ indices ($\Kb \geq \Ka$) as the effective output from the CS decoder.


\subsubsection{Message Passing for Gaussian MAC}
\label{sec:BP_GMAC}
The compressed sensing decoder outputs a set of $\Kb$ interleavers and corresponding fading coefficient estimates, which are used as input by the message passing decoder. The channel coding part of the received signal can be expressed as
\begin{align*}
\yv_{\mathrm{c}} &\coleq \yv[\Np+1:N_{\mathrm{t}}] = \sum_{k=1}^{\Kb}\hat{h}_k\pi_{{w_{\mathrm{p}}^{k}}}(\vv''_k) + \zv[\Np+1:N_{\mathrm{t}}]\\
&=\sum_{k=1}^{\Ka}\hat{h}_k\pi_{{w_{\mathrm{p}}^{k}}}(\vv''_k)+\sum_{k=\Ka+1}^{\Kb}\hat{h}_k\pi_{{w_{\mathrm{p}}^{k}}}(\underline{0}) + \zv[\Np+1:N_{\mathrm{t}}].
\end{align*}
Note that the received signal includes contributions from interleavers that were not employed by any of the $\Ka$ active users.
The $\Kb-\Ka$ additional interleavers can be viewed as the ones employed by fictitious users, each of them transmitting a zero signal.

\begin{figure}[h!]
  \centering
  \resizebox{1\textwidth}{!}{\tikzstyle{cnode}=[circle,draw]
\tikzstyle{cgnode}=[circle,draw]
\tikzstyle{cnode}=[circle,draw]
\tikzstyle{crnode}=[circle,draw]
\tikzstyle{conode}=[rectangle,draw]
\tikzstyle{cpnode}=[circle,draw]
\tikzstyle{rnode}=[rectangle,draw,outer sep=0pt]
\tikzstyle{prnode}=[rectangle,rounded corners,fill=blue!50,text width=4.5em,text centered,outer sep=0pt]
\tikzstyle{prnodebig}=[rectangle,rounded corners,fill=blue!50,text width=7em,text centered,outer sep=0pt]
\tikzstyle{prnodesimple}=[rectangle,draw,text width=4.5em,text centered,outer sep=0pt]
\tikzstyle{bigsnake}=[fill=green!50,snake=snake,segment amplitude=4mm, segment length=4mm, line after snake=1mm]
\tikzstyle{smallsnake}=[snake=snake,segment amplitude=0.7mm, segment length=4mm, line after snake=1mm]

\begin{tikzpicture}[every node/.style={scale=0.9}]
\begin{scope}[node distance=2cm,>=angle 90,semithick]
\node[cnode] (v1) {$v_1$};
\node[crnode] (v2)[right of=v1,xshift=-1] {$v_2$};
\node[cgnode] (v3)[right of=v2,xshift=-1] {$v_3$};
\node[conode] (c1)[above of=v2,yshift=-0.cm,xshift=-1cm] {$c_1$}
  edge[black,line width=1.3pt,decoration={markings,mark=at position 0.5 with {\arrow{>}}},postaction={decorate}] node[near end,above,xshift=-0.4cm]{$m^1_{c \rightarrow v}$}(v1)
  edge[black,line width=1.3pt,decoration={markings,mark=at position 0.5 with {\arrow{<}}},postaction={decorate}] node[near end,left,xshift=0.1cm]{$m^1_{v \rightarrow c}$} (v2);
  \node[conode] (c2)[above of=v2,xshift=1cm,yshift=-0 cm] {$c_2$}
  edge[black,line width=1.3pt] node[very near end,above]{} (v2)
  edge[black,line width=1.3pt] node[very near start,right]{}(v3);
  \node (user1)[above of =c1,xshift=1 cm,yshift=-1.5cm]{\large{User 1}};
\draw[black,line width=1.3pt] (v3) -- node[very near start,left]{}(c1.295);

\node[cgnode] (v4)[right of=v3,xshift=0] {$v_1$};
\node[crnode] (v5)[right of=v4,xshift=-1] {$v_2$};
\node[cgnode] (v6)[right of=v5,xshift=-1] {$v_3$};
\node[conode] (c3)[above of=v5,yshift=-0.cm,xshift=-1cm] {$c_1$}
  edge[black,line width=1.3pt,decoration={markings,mark=at position 0.5 with {\arrow{>}}},postaction={decorate}] node[near end,above,xshift=-0.4cm]{$m^2_{c \rightarrow v}$}(v4)
  edge[black,,line width=1.3pt,decoration={markings,mark=at position 0.5 with {\arrow{<}}},postaction={decorate}] node[near end,left,xshift=0.1cm]{$m^2_{v \rightarrow c}$} (v5);
\node[conode] (c4)[above of=v5,xshift=1cm,yshift=-0 cm] {$c_2$}
  edge[black,line width=1.3pt] node[very near end,above]{} (v5)
  edge[black,line width=1.3pt] node[very near start,right]{}(v6);  
\draw[black,line width=1.3pt] (v6) -- node[very near start,left]{}(c4.295);
\draw[black,line width=1.3pt] (v6) -- node[very near start,left]{}(c3.315);
\node (user2)[above of =c3,xshift=1 cm,yshift=-1.5cm]{\large{User 2}};
\node (vnode)[left of =v1,xshift= -0.3 cm,yshift=0 cm]{\large{Variable nodes}};
\node (vnode)[left of =c1,xshift= -1.3 cm,yshift=0 cm]{\large{Check nodes}};

\node[conode] (m1)[below of=v1,xshift=-0 cm]{+};
\node[conode] (m2)[right of=m1,xshift= 0 cm]{+};
\node[conode] (m3)[right of=m2,xshift= 0 cm]{+};
\node[conode] (m4)[below of=v4,xshift=- 0 cm]{+};
\node[conode] (m5)[right of=m4,xshift=-0 cm]{+};
\node[conode] (m6)[right of=m5,xshift= 0 cm]{+};
\node (vnode)[left of =m1,xshift= -0.4 cm,yshift=0 cm]{\large{MAC nodes}};
\draw[black,line width=1.3pt,decoration={markings,mark=at position 0.5 with {\arrow{>}}},postaction={decorate}] (m1) -- node[near start,left]{$m^1_{+ \rightarrow v}$}(v1);
\draw[black,line width=1.3pt] (m1) -- node[very near end,below,xshift=0.1cm]{}(v4);
\draw[black,line width=1.3pt,decoration={markings,mark=at position 0.5 with {\arrow{<}}},postaction={decorate}] (m2) -- node[very near end,left,xshift=-0.1cm]{$m^1_{v \rightarrow +}$}(v2);
\draw[black,line width=1.3pt,decoration={markings,mark=at position 0.5 with {\arrow{>}}},postaction={decorate}] (m6) -- node[midway,right,xshift=-0.1cm]{$m^2_{+ \rightarrow v}$}(v6);
\draw[black,line width=1.3pt] (m3) -- node[very near end,left,xshift=-0.1cm]{}(v3);
\draw[black,line width=1.3pt] (m3) -- node[very near end,left,xshift=-0.1cm]{}(v6);
\draw[black,line width=1.3pt] (m2) -- node[very near end,left,xshift=-0.1cm]{}(v5);
\draw[black,line width=1.3pt] (m4) -- node[very near end,left,xshift=-0.1cm]{}(v4);
\draw[black,line width=1.3pt,decoration={markings,mark=at position 0.3 with {\arrow{<}}},postaction={decorate}] (m5) -- node[midway,right,xshift=-0.05cm]{$m^2_{v \rightarrow +}$}(v5);

\end{scope}    
\end{tikzpicture}}
  \caption{ Tanner graph representation of the coding scheme}
  \label{fig:BP_computationgraph}
\end{figure}
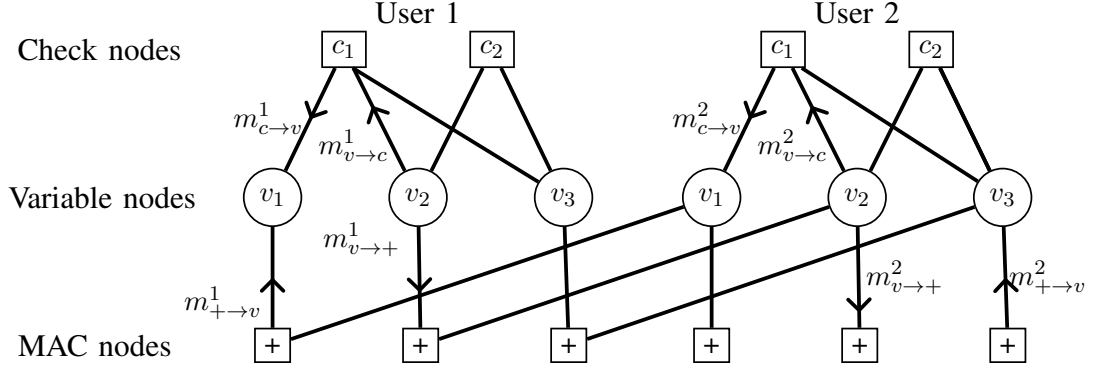

For ease of exposition, we describe the message passing rules for $\Kb=\Ka=2$ and $\hat{h}_1=\hat{h}_2=1$.
It can be generalized to larger values of $\Kb$, $\Ka$ in a straightforward manner.
Given the received signal $\yv_{\mathrm{c}}$ the joint BP decoder proceeds iteratively passing messages along the edges of a Tanner graph that represents the coding scheme. Such a Tanner graph and the associated messages that are passed during the decoding are shown in Fig.~\ref{fig:BP_computationgraph}. The nodes marked $v$, $c$ and $+$ represent variable nodes, check nodes and MAC nodes, respectively.
Throughout this section, we use superscript to distinguish between users $1$ and $2$. The following messages are passed at every iteration along the edges of the Tanner graph.
\begin{itemize}
\item $m^{1}_{v \rightarrow c}(e)$: Messages passed from bit node to check node along edge $e$ of user 1.
\item $m^{1}_{c \rightarrow v}(e)$: Messages passed from variable node to check node along edge $e$ of user 1.

\item $m^{1}_{v \rightarrow +}(e)$: Message passed from variable node of user~$1$ to MAC node along edge $e$.
\item $m^{1}_{+ \rightarrow v}(e)$: Message passed from  MAC node to variable node of user~$1$ along edge $e$.
\end{itemize}
The messages for user~$2$ are defined similarly. The rules for message passing are somewhat standard.

Given an edge $e$ between a variable node and a check node, let $v_e$ and $c_e$ denote the variable node and check node connected to $e$, respectively. Similarly, given an edge $e$ between a variable node and a MAC node, let $v_e$ and $+_e$ denote the variable node and MAC node connected to $e$, respectively.
Let $\mathcal{N}(c_e)$ be the set of edges connected to check node $c_e$,
and $\mathcal{N}_c(v_e)$ represent the set of edges that connect variable node $v_e$ to check nodes.
Let $\mathcal{N}_+(v_e)$ denote the set of edges that connect the variable node $v_e$ to MAC nodes.
Let $\mathcal{N}(+_e)$ be the set of edges connected to MAC node $+_e$. 

At the bit node, we have
\begin{align*}
m^{1}_{v \rightarrow c}(e)&= \sum_{f \in \mathcal{N}_+(v_e) }m^{1}_{+ \rightarrow v}(f) + \sum_{e_i \in \mathcal{N}_c(v_e)\setminus e} m^{1}_{c \rightarrow v}(e_i)\\
m^{1}_{v \rightarrow +}(e)&=\sum_{e_i\in \mathcal{N}_c(v_e)} m^{1}_{c \rightarrow v}(e_i) + \sum_{f \in \mathcal{N}_+(v_e)} m^1_{+ \rightarrow v}(f).
\end{align*}
The LDPC check nodes implement
\begin{align*}
m^{1}_{c \rightarrow v}(e)&=2~\tanh^{-1}\left( \prod_{e_i\in \mathcal{N}(c_{e})\setminus e} \tanh\left(\frac{m^{1}_{v \rightarrow c}(e_i)}{2}\right)\right) .
\end{align*}
As discussed earlier, the receiver sees  a $\frac{1}{\Nc} \sum_{l=1}^{L} \nu_l Nl$-user Gaussian MAC because of the sparse nature  of transmissions, which enables the receiver to do optimal demodulation at MAC nodes.
The message at the MAC node corresponding to the $j$th use of the channel is updated by
\begin{equation} \label{Eqn:MACnodeBP}
\begin{split}
m^{1}_{+ \rightarrow v}(e) = h(m^{2}_{v \rightarrow +}(f),\underline{y}_{c}(j)), \\
\end{split}
\end{equation}
where $f \in \mathcal{N}(+_{e})\setminus e$ is the neighboring edge of $e$ at a MAC node and
$h(\ell,y;P) =\log \frac{1+e^{l}e^{2(y-\sqrt{P})}}{e^{l}+e^{-2(y+\sqrt{P})}}$.
The function $h(\ell,y;P)$ can be viewed as the log-likelihood of variable $x_2$ when $y=x_1+x_2+z$, $x_1, x_2\in\{\pm \sqrt{P}\}$, the log-likelihood ratio of variable $x_1$ is known to be $\ell$, and $z\sim \mc{N}(0,1)$.  

\section{Density evolution and Code Construction}
\label{sec:code_construction}
\begin{figure}
    \centering
    \tikzstyle{cnode}=[circle,draw]
\tikzstyle{cgnode}=[circle,draw]
\tikzstyle{cnode}=[circle,draw]
\tikzstyle{crnode}=[circle,draw]
\tikzstyle{conode}=[rectangle,draw]
\tikzstyle{cpnode}=[circle,draw]
\tikzstyle{rnode}=[rectangle,draw,outer sep=0pt]
\tikzstyle{prnode}=[rectangle,rounded corners,fill=blue!50,text width=4.5em,text centered,outer sep=0pt]
\tikzstyle{prnodebig}=[rectangle,rounded corners,fill=blue!50,text width=7em,text centered,outer sep=0pt]
\tikzstyle{prnodesimple}=[rectangle,draw,text width=4.5em,text centered,outer sep=0pt]
\tikzstyle{bigsnake}=[fill=green!50,snake=snake,segment amplitude=4mm, segment length=4mm, line after snake=1mm]
\tikzstyle{smallsnake}=[snake=snake,segment amplitude=0.7mm, segment length=4mm, line after snake=1mm]

\begin{tikzpicture}[every node/.style={scale=1.8}]
\begin{scope}[node distance=2cm,>=angle 90,semithick]
\node[cnode] (v1) {$v_1$};
\node[crnode] (v2)[right of=v1] {$v_2$};
\node[cgnode] (v3)[right of=v2] {$v_3$};
\node[cpnode] (v4)[right of=v3] {$v_4$};
\node[conode] (c1)[below of=v2,yshift=-0cm,xshift=-0.5cm] {$c_1$}
  edge[black] node[near end,above]{$e_1$}(v1)
  edge[black] node[left]{$e_3$} (v2);
\draw[black] (v4.210) -- node[very near start,left]{$e_7$}(c1.60);
\draw[black] (v4.225) -- node[near start,below]{$e_8$}(c1.35);
\node[conode] (c2)[below of=v3,xshift=0.5cm,yshift=-0 cm] {$c_2$}
  edge[black] node[very near end,above]{$e_2$} (v1)
  edge[black] node[very near start,right]{$e_9$}(v4);
\draw[black] (v3) --node[very near start,left]{$e_5$} (c1.70);
\draw[black] (v3.280) --node[very near start,left]{$e_6$} (c2.90);
\draw[black] (v2) --node[very near start,above]{$e_4$}(c2);
\end{scope}    
\end{tikzpicture}
    \caption{Example of a protograph}
    \label{fig:protex}
\end{figure}
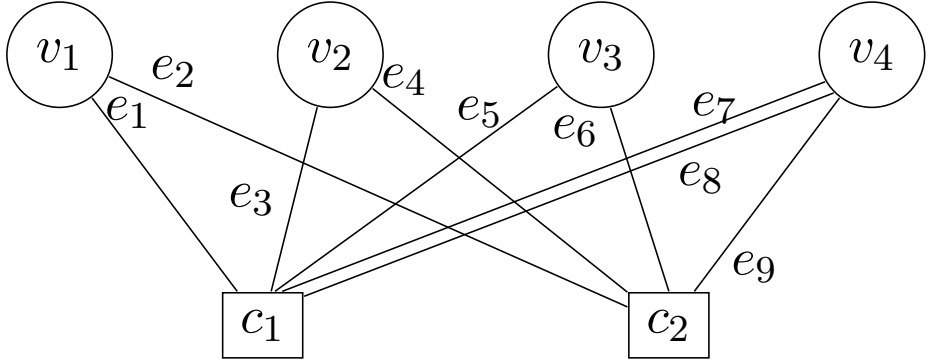

A protograph $\mathcal{G}=(V\cup C,\mathcal{E})$ is a
bipartite graph with the bipartition $V$ and $C$ called the set of
variable and check nodes, respectively.
The set $\mathcal{E}$ of undirected edges specifies the connections between variable nodes in $V$ and check nodes in $C$.
The $i$th variable node, check node and edge in the protograph are denoted, respectively, by $v_i$, $c_i$ and $e_i$. An example of a protograph appears in Fig.~\ref{fig:protex}.
An LDPC code can be obtained from the protograph  by copy-and-permute operation.
Since the codes obtained form a multi-edge-type ensemble with $|\mathcal{E}|$ edge types, density evolution proceeds with $ |\mathcal{E}|$ types of messages, one for each edge in the protograph~\cite{mct}. 

Let $\nu(x) \coleq \sum_{l=1}^L \nu_lx^l$ denote the repetition degree distribution (d.d.), where $\nu_l$ represents the fraction of active users who repeat their codewords $l$ times. This structure induces a degree distribution on the MAC nodes given by
$G(x) \coleq \sum_{i=1}^{L} G_i x^i$, where $G_i$ is the fraction of time instants where $i$ users transmit. When the interleavers in \eqref{eqn:codeconstruction1} are chosen uniformly at random from the set of all possible interleavers of length $N_c$, it can be seen that $G_i={N_c \choose i} q^i (1-q)^{N_c-i}$ with $q=\frac{1}{{N_c}}\sum_{l=1}^{L}\nu_lNl$.
In the limit as $N_c$ grows large, $G(x)$ converges to $\sum_{i=0}^\infty \frac{e^{-q} (q x)^i}{i!}$. The edge perspective MAC node degree distribution, denoted by $\gamma(x)$, is given by $G'(x)/G'(1)$.


Next, we introduce notations required to describe the density evolution (DE). Without loss of generality, we consider a coded bit whose value is +1. Under the assumption that messages (log-likelihood ratios) along edges are Gaussian with mean $\sigma^2/2$ and variance $\sigma^2$, the mutual information (MI) between the message along an edge and the codeword bit associated with it is given by $J(\sigma)$ \cite{Liva}  
\begin{align*}
    J(\sigma)=1-\int_{-\infty}^{+\infty}{1\over \sqrt{2\pi\sigma^{2}}}e^{-{(y-\sigma^{2}/2)^{2}\over 2\sigma^{2}}}\cdot\log_{2}(1+e^{-y})dy.
\end{align*}
Note that $(J^{-1}(I))^2$ is the variance of the LLRs when the MI between the message and the corresponding variable is $I$.

Consider the messages passed along the edges during the $t$th iteration for a user who repeats its bits $l$ times.
Let $I_{v\rightarrow c}^t(e_i,l)$ represent the MI between the message from variable node to check node along the edge type $e_i$ and the associated codeword bit.
Similarly, define $I_{c \rightarrow v}^t(e_i,l)$  as the MI between the message along the edge type $e_i$ from check node to variable node and the associated codeword bit. 
Let $I_{v\rightarrow +}^t(v_i,l)$ denotes the MI between the message from variable node $v_i$ to the  MAC node and the codeword bit associated with $v_i$.
Let $I_{+\rightarrow v}^t$ denote the average MI between the message from MAC node to variable node and the associated codeword bit.
Let $I_{v\rightarrow +}^t$ denote the average MI between the message from variable nodes to MAC nodes and the codeword bits.
Finally, let $I_{\mathrm{APP}}^t(v,l)$ denote the mutual information between the posterior log-likelihood-ratio (LLR) evaluated at variable node $v$ and the associated codeword bit.

Consider a MAC node with two users and BPSK modulation without additive noise. Let $\sigma^2$ be the variance of a priori (incoming) LLRs at the MAC node. We assume that a MAC node performs soft interference cancellation and that the remaining interference at the MAC node is Gaussian. Let $\phi(\sigma)$ denote the minimum mean squared error in the estimate of a variable after soft interference cancellation. Then, $\phi(\sigma)$ is given by \cite{verdu2005}
\begin{align*}
    \phi \left( \sigma \right) = 1 - \int_{ - \infty }^\infty {\frac{{{e^{ - \frac{{{y^2}}}{2}}}}}{{\sqrt {2\pi } }}\tanh \left( {\frac{\sigma^2 }{4} - \frac{\sigma }{2} y} \right)dy}.
\end{align*}

For a user whose codeword is repeated $l$ times, we start the density evolution recursion by initializing $I_{v \rightarrow +}^{0}$ to zero. Then,
\begin{align*}
    \label{Eqn:MAC_update}
    I_{+ \rightarrow v}^t=\sum_k \gamma_k J\left(\frac{2}{\sqrt{\sigma^2_n+(\sigma^t_{I,k})^2}}\right), \nonumber
\end{align*}
where $(\sigma^t_{I,k})^2$ is given by \cite{verdu2005},
\begin{align*}
    (\sigma^t_{I,k})^2=(k-1)\phi \left(J^{-1}(I_{v \rightarrow +}^{t-1})\right).
\end{align*}
We also have
\begin{equation*}
\begin{split}
    \label{Eqn:variable_update}
    &I_{v \rightarrow c }^t(e,l)=\\&J\left(\sqrt{\sum_{e_i \in \mathcal{N}_c(v_{e})\setminus e}[J^{-1}(I_{c \rightarrow v}^{t}(e_i,l))]^2+l[J^{-1}(I_{+ \rightarrow v}^t)]^2}\right)
\end{split}
\end{equation*}
\begin{equation*}
\begin{split}
    \label{Eqn:check_update}
    I_{c \rightarrow v}^t(e,l)=1-J\left(\sqrt{\sum_{e_i\in \mathcal{N}(c_{e})\setminus e}[J^{-1}(1-I_{v \rightarrow c}^{t}(e_i,l))]^2}\right)
    \end{split}
\end{equation*}
\begin{equation*}
\begin{split}
    \label{Eqn:rep_MAC_update}
    &I_{v \rightarrow +}^t(v_i,l)=\nonumber\\&J\left(\sqrt{(l-1)[J^{-1}(I_{+ \rightarrow  v_i}^{t})]^2+\sum_{e\in \mathcal{N}_v(v_i)}[J^{-1}(I_{c \rightarrow v }^{t}(e,l))]^2}\right).
    \end{split}
\end{equation*}
\begin{align*}
    I_{v \rightarrow +}^t(l)=\frac{1}{|V|}\sum_i I_{v\rightarrow +}^t(v_i,l),
\end{align*}
where $|V|$ is the number the number of variable nodes in the protograph.
Finally,
\begin{align*}
    \label{Eqn:rep_mac_avg_update}
\textstyle    I_{v \rightarrow +}^t=\sum_{l=1}^{L}\nu_l I_{v \rightarrow +}^t(l)
\end{align*}
\begin{equation*} \label{Eqn:decision}
\begin{split}
    &I_{\mathrm{APP}}(v_i,l) \\
    &=J\left(\sqrt{\sum_{e\in \mathcal{N}_v({v_i})}[J^{-1}(I_{c \rightarrow v}^{t}(e))]^2+[J^{-1}(I_{v \rightarrow +}^t(v_i,l))]^2}\right) .
\end{split}
\end{equation*}
The density evolution threshold is defined as the minimum $E_b/N_0$ for 
which $I_{\mathrm{APP}} (v_i,u_l) \rightarrow 1$, as $t \rightarrow \infty $, for all $v_i$ and $l \in \{1,2, \ldots, L\}$. 

We use differential evolution \cite{Storn} to optimize the protographs and $\nu(x)$ by using the density evolution threshold as the cost function.
We lift optimized protographs to codes using the progressive edge growth algorithm. Even though DE thresholds are meaningful benchmarks only for asymptotic lengths.
Nevertheless, designing codes based on DE thresholds offers a principled way to optimize the performance of our system. Simulation results show that this approach is efficient even for short block lengths.

\section{Numerical Results}
The parameters we select for our numerical study are: (i) Number of bits each user intends to transmit $B=100$, (ii) Total number of channel uses $N_{\mathrm{t}}=30000$, (iii) Total number of active users $\Ka \in [25:300]$, (iv) Maximum per user error probability $P_e \leq \epsilon=0.05$. These value are chosen to match the parameters employed in \cite{ordentlich2017low} for ease of comparison.

We fix $\Bp=15$ and $\Np=2000$. The sensing matrix for the CS encoder is constructed as follows. We pick $\Np/2$ rows uniformly at random from the discrete Fourier transform (DFT) matrix of dimension $M_{\mathrm{p}}$. The real and imaginary parts of each row are then stacked to form a $\Np \times M_{\mathrm{p}}$ real sensing matrix $\mathbf{A}$; entries are normalized to meet the power constraint.
Matrices constructed this way satisfy restricted isometry property (RIP) with high probability and are a good choice for the sensing matrix in noisy compressed sensing~\cite{haviv2017restricted}. This then yields the parameters for the channel coding part, with $\Bc=85$ and $\Nc=28000$.

For a fixed value of $\Ka$, computing the required SNR involves solving the optimization problem
\begin{align}
&\frac{E_b}{N_0} = \min_{P_1,P_2,K_b} \frac{\Np P_1+\Nc P_2}{2 B} \label{eqn:SNRdefinition} \\
&\text{such that }~ \Pr(\mc{E}|P_1,P_2,\Kb) \leq \epsilon. \nonumber
\end{align}
\vspace{-0.1in}
\begin{table}[ht]
  \centering
  \begin{tabular}{|c|c|c|c|}
    \hline
    \bf{Number of Users}&$25-125$&$150-200$&$225-300$ \\
    \hline
    \bf{Rate}& $0.125$  & $0.25$ &  $0.4$  \\
    \hline
  \end{tabular}
  \caption{Code rates corresponding to number of active users}
  \label{tab:code_param}
\end{table}

 The proposed scheme is evaluated as follows.
 For each $\Ka \in \{25,50,\ldots,300\}$, we use the optimization procedure described in Section~\ref{sec:code_construction} to optimize the protograph for the LDPC code and the repetition d.d. $\nu(x)$. The function $g(w_{\rm p})$ is then chosen to induce this degree distribution.
 Although one needs to solve the optimization problem in \eqref{eqn:SNRdefinition} to achieve the optimal SNR, this task is computationally complex due to the parameter space being huge.
 Alternatively, using simulations, we found $\Kb=110$ to be a suitable choice when $\Ka=100$, and thus we fix $\Kb=\lceil 1.1\Ka\rceil$.
 With $\Kb$ fixed, we sweep over all possible combinations of $P_1$, $P_2$ in a two-dimensional grid of SNR values, with a resolution of 0.5~dB in each dimension, for the compressed sensing and the channel coding components.
 We emphasize that this only results in an approximate solution to the above optimization problem.
 \begin{figure}[t]
    \centering
    \begin{tikzpicture}
\definecolor{mycolor1}{rgb}{0.63529,0.07843,0.18431}%
\definecolor{mycolor2}{rgb}{0.00000,0.44706,0.74118}%
\definecolor{mycolor3}{rgb}{0.00000,0.49804,0.00000}%
\definecolor{mycolor4}{rgb}{0.87059,0.49020,0.00000}%
\definecolor{mycolor5}{rgb}{0.00000,0.44700,0.74100}%
\definecolor{mycolor6}{rgb}{0.74902,0.00000,0.74902}%

\begin{axis}[%
font=\small,
width=7cm,
height=6cm,
scale only axis,
every outer x axis line/.append style={white!15!black},
every x tick label/.append style={font=\color{white!15!black}},
xmin=25,
xmax=300,
xtick = {25,50,100,...,300},
xlabel={Number of active users $\Ka$},
xmajorgrids,
every outer y axis line/.append style={white!15!black},
every y tick label/.append style={font=\color{white!15!black}},
ymin=1,
ymax=5,
ytick = {1,...,5},
ylabel={Required $E_b/N_0$ (dB)},
ymajorgrids,
legend style={at={(0,1)},anchor=north west, draw=black,fill=white,legend cell align=left}
]

\addplot [color=black,solid,line width=2.0pt,mark size=1.4pt,mark=square,mark options={solid}]
  table[row sep=crcr]{
 25	2\\
50	2.1\\
75	2.2\\
100	2.41\\
125 2.57\\
150 4\\
};
\addlegendentry{Rate=$0.125$, $\nu(x)=x^2$};


\addplot [color=mycolor2,densely dotted,line width=2.0pt,mark size=1.4pt,mark=square,mark options={solid}]
  table[row sep=crcr]{25	2.75\\
50	2.75\\
100	3.19\\
125	3.24\\
150	3.3\\
200	3.8\\
250 4.36\\
300 5.35\\
};
\addlegendentry{Rate=$0.4$, $\nu(x)=x^2$};


%

\end{axis}



\end{tikzpicture}
    \caption{Minimum $E_b/N_0$ required as a function of number of users and codes used by each user.}
    \label{fig:different_rate_comparison}
\end{figure}
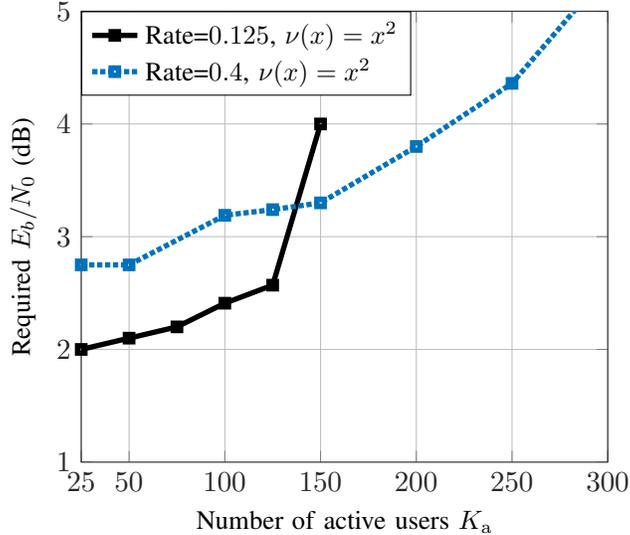

 
 We perform full-blown simulations of the proposed scheme to account for preamble collision and missed detection of the CS decoder.
 The former event forces colliding users to pick the same parameters (interleaver and code rate) for the channel code; while the latter fails to provide information regarding the parameters of the channel code chosen by the active users whose preambles are not detected by the CS decoder.
 By choosing $P_1$ and $P_2$ as the approximate solutions to optimization problem \eqref{eqn:SNRdefinition}, we ensure that these events occur with a low probability.
 \subsection{Selection of codes as a function of number of users }
 
The rate of the protograph LDPC code has a significant effect on the required $E_b/N_0$ for a fixed value of $\Ka$. 
For different rates, the minimum $E_b/N_0$ required to achieve a probability error of $0.05$ is plotted in Fig.~\ref{fig:different_rate_comparison} as a function of the number of users. It can be seen that
the optimal rate changes with the number of active users $\Ka$. For example, for $P_e=0.05$ and $\Ka =125$, an $E_b/N_0$ of $2.47$~dB is required if a rate-$0.125$ LDPC code with $\nu(x)=x^2$ is utilized, whereas an $E_b/N_0$ of $3.24$~dB is required if a rate-$0.4$ LDPC code with the same repetition pattern is employed. For a fixed number of users, we choose the rate through simulations to minimize the $E_b/N_0$ required to achieve a target probability of error.

\subsection{Comparison with existing schemes}
 \begin{figure}[t]
\centering
  \begin{tikzpicture}
\definecolor{mycolor1}{rgb}{0.63529,0.07843,0.18431}%
\definecolor{mycolor2}{rgb}{0.00000,0.44706,0.74118}%
\definecolor{mycolor3}{rgb}{0.00000,0.49804,0.00000}%
\definecolor{mycolor4}{rgb}{0.87059,0.49020,0.00000}%
\definecolor{mycolor5}{rgb}{0.00000,0.44700,0.74100}%
\definecolor{mycolor6}{rgb}{0.74902,0.00000,0.74902}%

\begin{axis}[%
font=\footnotesize,
width=7cm,
height=6cm,
scale only axis,
xmin=25,
xmax=275,
xtick = {25,75,...,275},
xlabel={Number of active users $K$},
xmajorgrids,
ymin=0,
ymax=12,
ytick = {0,2,...,12},
ylabel={Required $E_b/N_0$ (dB)},
ylabel near ticks,
ymajorgrids,
legend style={font=\tiny, at={(1.7,1)},anchor=north east, draw=black,fill=white,legend cell align=left}
]

\addplot [color=black,dotted,line width=1.5pt]
  table[row sep=crcr]{
 25	0.25\\
50	0.3\\
75	0.35\\
100	0.4\\
125	0.45\\
150	0.5\\
175	0.55\\
200	0.6\\
225	0.95\\
250	1.25\\
275	1.55\\
300	1.8\\
};
\addlegendentry{Random Coding \cite{polyanskiy2017perspective}};

\addplot [color=mycolor1,dotted,line width=1.5pt]
  table[row sep=crcr]{25	2.26\\
50	2.88\\
75	3.9\\
100	5.03\\
125	5.8798\\
150	7.3954\\
175	8.6199\\
200	9.7328\\
225	11.1761\\
250	12.6127\\
275	13.3907\\
300	14.9116\\
};
\addlegendentry{4-Fold ALOHA \cite{ordentlich2017low}};

\addplot [color=mycolor2,solid,line width=1.5pt]
  table[row sep=crcr]{25	3.18\\
50	3.52\\
75	4.64\\
100	5.61\\
125	5.85\\
150	6.46\\
175	6.72\\
200	7.41\\
225	7.6772\\
250	8.3217\\
275	8.8428\\
300	9.352\\
};
\addlegendentry{SIC T=4, \cite{vem2017user}};

\addplot [color=mycolor4,solid,line width=1.5pt,mark size=2.0pt,mark=triangle,mark options={solid,rotate=90}]
  table[row sep=crcr]{
  25  3.15\\
50	3.2\\
75	3.3\\
100	3.6\\
125	4\\
150	4.85\\
175	5.23\\
200	5.52\\
225	6.05\\
250	6.8\\
275	7.4\\
300	8.22\\
};
\addlegendentry{CCS \cite{amalladinne2018coupled}};

\addplot [color=black,solid,line width=1.5pt]
  table[row sep=crcr]{
25	3\\
50	3.5\\
75  3.5\\
100	4\\
125	4\\
150 4.5\\
175 5\\
200	5\\
225 5.5
250	5.8\\
275 5.8\\
300	6\\
};
\addlegendentry{AMP+Tree Code, \cite{Giuseppe}};

\addplot [color=mycolor3,solid,line width=1.5pt,mark size=1.0pt,mark=square,mark options={solid}]
  table[row sep=crcr]{
  25  2\\
50	2.1\\
75	2.2\\
100	2.41\\
125	2.57\\
150	2.81\\
175	3\\
200 3.4\\
225 3.88\\
250 4.36\\
275 4.87\\
300 5.35\\
};\addlegendentry{Proposed Scheme};

\addplot [color=cyan,solid,line width=2.0pt,mark size=1.4pt,mark=triangle,mark options={solid}]
  table[row sep=crcr]{
  25  0.73\\
50	1.15\\
75	1.63\\
100	2.06\\
125	2.48\\
150	2.99\\
175	3.43\\
200 3.89\\
225 4.41\\
250 4.91\\
275 5.26\\
300 5.49\\
};
\addlegendentry{IRSA + Polar Coding \cite{marshakov2019polar}};
\addplot [color=mycolor1,solid,mark=square,line width=2.0pt]
  table[row sep=crcr]{
   2     0.3\\
  10    0.5\\
 25	0.55\\
50	0.6\\
75 0.7\\
100 0.75\\
125 1.15\\
150 1.5\\
175 2\\
200 2.7\\
225 3.5\\
250 4.3\\};
\addlegendentry{Spreading+Polar Coding \cite{pradhan2019polar} };

\node[] at (axis cs: 300,5.15) {\scriptsize \textcolor{red}{O}};
\node[] at (axis cs: 275,4.69) {\scriptsize \textcolor{red}{O}};
\node[] at (axis cs: 225,3.7) {\scriptsize \textcolor{red}{O}};
\end{axis}
\end{tikzpicture}%
  \caption{Minimum $E_b/N_0$ required vs. number of users.}
  
  \label{fig:simulationresults30000}
\end{figure}
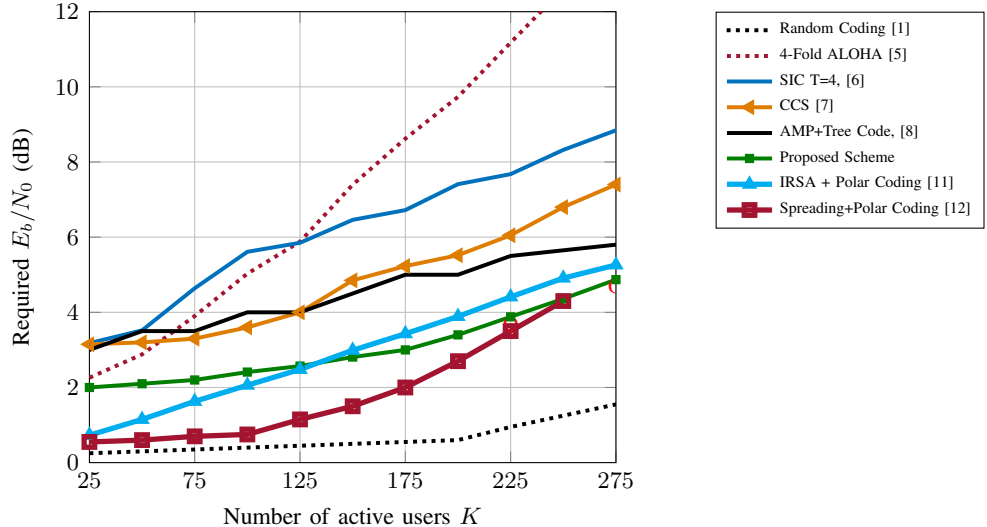

The performance of the scheme developed herein is compared to the existing schemes in Fig.~\ref{fig:simulationresults30000}. Rates of LDPC codes used for each value of $\Ka$ are given in Table~\ref{tab:code_param}. 
 In the simulation of proposed scheme, repetition pattern $x^2$ is used for all values of $\Ka$. 
 The obtained simulation results show that the proposed scheme performs better than the schemes in \cite{vem2017user,amalladinne2018coupled,Giuseppe,calderbank2018chirrup}. For example, at $\Ka=175$, the proposed scheme outperforms the scheme in \cite{Giuseppe} by $1.5$~dB.
 It can also be seen that the scheme in \cite{pradhan2019polar} outperforms the proposed scheme when $\Ka \leq 225$.
However, the polar coding based approach in \cite{pradhan2019polar} does not scale well with the number of active users.
It is pertinent to note that performance of the proposed scheme can be further improved by using irregular repetition patterns across users. 
  \begin{table}[h!]
  \centering
  \begin{tabular}{|c|c|c|}
    \hline
    \bf{Number of Users} & \bf{Repetition pattern $\nu(x)$} & \bf{Rate}\\
    \hline
    $225$ & $0.12x+0.88x^2$ & $0.4$\\
    \hline
    $275$ & $0.18x+0.82x^2$ & $0.4$\\
    \hline
    $300$ & $0.18x+0.82x^2$ & $0.4$\\
    \hline
  \end{tabular}
  \caption{Repetition patterns and code rates corresponding to number of active users.}
  \label{tab:rep_pattern}
\end{table}

In Fig.~\ref{fig:simulationresults30000}, the red circles indicate the $E_b/N_0$ required when the optimized repetition d.d. given in Table~\ref{tab:rep_pattern} is used. A small improvement of about 0.2~dB results from using an irregular repetition d.d.
 \subsection{Rayleigh fading channel}
 
\begin{figure}
\centering
  \begin{tikzpicture}{scale=1}
\definecolor{mycolor1}{rgb}{0.63529,0.07843,0.18431}%
\definecolor{mycolor2}{rgb}{0.00000,0.44706,0.74118}%
\definecolor{mycolor3}{rgb}{0.00000,0.49804,0.00000}%
\definecolor{mycolor4}{rgb}{0.87059,0.49020,0.00000}%
\definecolor{mycolor5}{rgb}{0.00000,0.44700,0.74100}%
\definecolor{mycolor6}{rgb}{0.74902,0.00000,0.74902}%

\begin{axis}[%
font=\small,
width=7cm,
height=6cm,
scale only axis,
every outer x axis line/.append style={white!15!black},
every x tick label/.append style={font=\color{white!15!black}},
xmin=25,
xmax=250,
xtick = {25,50,100,...,250},
xlabel={Number of active users $\Ka$},
xmajorgrids,
every outer y axis line/.append style={white!15!black},
every y tick label/.append style={font=\color{white!15!black}},
ymin=7.5,
ymax=20,
ytick = {8,10,12,...,18},
ylabel={Required $E_b/N_0$ (dB)},
ylabel near ticks,
ymajorgrids,
legend style={at={(0,1)},anchor=north west, draw=black,fill=white,legend cell align=left}
]

\addplot [color=black,dotted,line width=2.0pt]
  table[row sep=crcr]{
 25	8.1\\
50	8.1\\
75	8.1\\
100	8.1\\
125	8.1\\
150	8.1\\
175	8.1\\
200	8.1\\
225	8.1\\
250	8.1\\
};
\addlegendentry{Optimal Decoder\cite{suhash2020fading}};

\addplot [color=mycolor3,solid,line width=2.0pt,mark size=1.4pt,mark=diamond]
  table[row sep=crcr]{25	10.6\\
50	10.6\\
75	10.65\\
100	10.7\\
125	11\\
150	11.25\\
175	11.5\\
200	11.75\\
225	12\\
250	12.25\\
};
\addlegendentry{4-fold ALOHA + FBL bound\cite{suhash2020fading}};


\addplot [color=mycolor2,solid,line width=2.0pt,mark size=1.4pt,mark=square,mark options={solid}]
  table[row sep=crcr]{25	10.6\\
50	10.65\\
75	10.65\\
100	10.65\\
125	10.95\\
150	11.4\\
175	12.8\\
200	14.2\\
225	16.4\\
250	19\\
};
\addlegendentry{4-fold ALOHA + LDPC\cite{suhash2020fading}};

\addplot [color=mycolor4,solid,line width=2.0pt,mark size=1.3pt,mark=triangle,mark options={solid,rotate=90}]
  table[row sep=crcr]{
  25  10.1\\
50	10.1\\
75	10.3\\
100	10.3\\
125	10.5\\
150	10.82\\
175	11.4\\
200	12\\
225	12.7\\
250	13.5\\
275	14.3\\
300	15.2\\
};
\addlegendentry{Proposed scheme};

\end{axis}

\end{tikzpicture}
  \caption{Minimum $E_b/N_0$ required vs. number of users for $\epsilon=0.1,B=100$ and $N_{\mathrm{t}}=30000$.}
  
  \label{fig:simulationresults_fading_30000}
\end{figure}
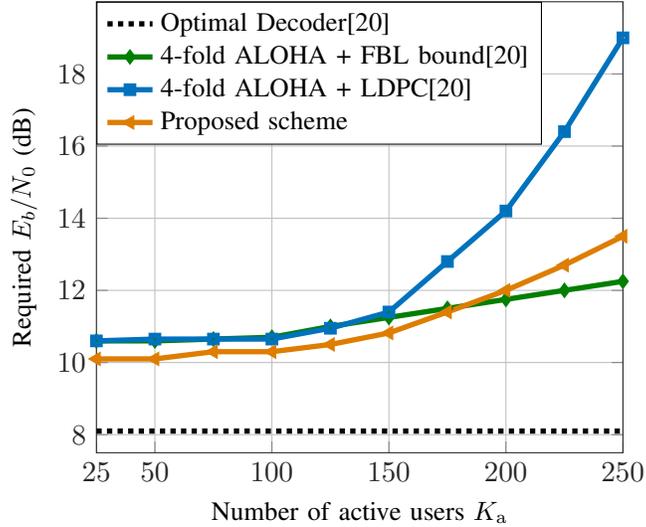

 The proposed scheme also performs well over Rayleigh fading channel. The performance of the scheme is compared to the existing schemes in Fig.~\ref{fig:simulationresults_fading_30000}.
  The simulation results show that the proposed scheme's performance is comparable to the scheme in \cite{suhash2020fading}, which assumes the existence of codes that achieve FBL capacity. 
 The proposed scheme outperforms the scheme in \cite{suhash2020fading} when it is simulated with LDPC codes in the regime $\Ka > 175$.
 
 \subsection{Comparison with IDMA}
\begin{figure}
    \centering
    \begin{tikzpicture}
\definecolor{mycolor1}{rgb}{0.63529,0.07843,0.18431}%
\definecolor{mycolor2}{rgb}{0.00000,0.44706,0.74118}%
\definecolor{mycolor3}{rgb}{0.00000,0.49804,0.00000}%
\definecolor{mycolor4}{rgb}{0.87059,0.49020,0.00000}%
\definecolor{mycolor5}{rgb}{0.00000,0.44700,0.74100}%
\definecolor{mycolor6}{rgb}{0.74902,0.00000,0.74902}%

\begin{axis}[%
font=\small,
width=7cm,
height=6 cm,
scale only axis,
every outer x axis line/.append style={white!15!black},
every x tick label/.append style={font=\color{white!15!black}},
xmin=5,
xmax=25,
xtick = {5,10,...,25},
xlabel={Number of active users $\Ka$},
xmajorgrids,
every outer y axis line/.append style={white!15!black},
every y tick label/.append style={font=\color{white!15!black}},
ymin=0,
ymax=15,
ytick = {0,5,...,15},
ylabel={Required $E_b/N_0$ (dB)},
ymajorgrids,
legend style={at={(0,1)},anchor=north west, draw=black,fill=white,legend cell align=left}
]

\addplot [color=black,dotted,line width=2.0pt]
  table[row sep=crcr]{
  5 0.2\\
 25	0.25\\
};
\addlegendentry{Random Coding\cite{polyanskiy2017perspective}};
\addplot [color=black,solid,line width=2.0pt,mark size=1.4pt,mark=square,mark options={solid}]
  table[row sep=crcr]{
 5	2.59\\
10	5.1\\
15	7\\
20	9\\
25 12\\
};
\addlegendentry{IDMA};


%

\end{axis}



\end{tikzpicture}%
    \caption{Minimum $E_b/N_0$ required  vs. no. of users for IDMA.}
    \label{fig:IDMA}
\end{figure}
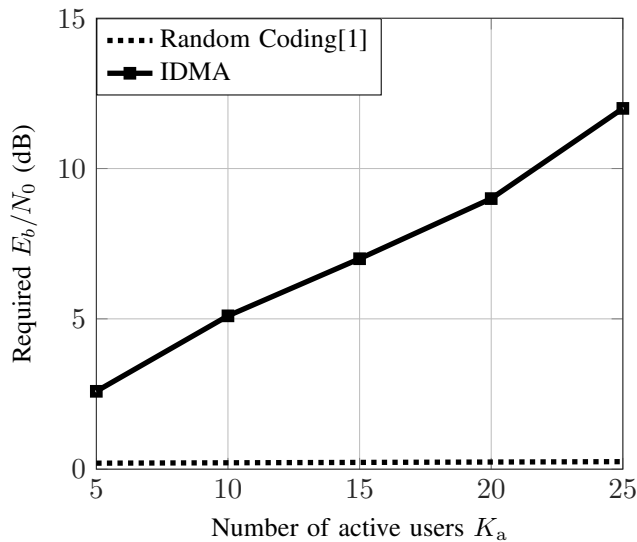

 We now present a comparison of the proposed scheme with conventional IDMA. Prior work has shown that IDMA is very effective when the number of users is small (less than 25-30) and block lengths are reasonably large \cite{ping2006interleave}, \cite{tenbrink}. 
 Designing very low rate ($approx$ 1/300) iteratively-decodable multi-user codes with short block lengths for a large number of users is a significant challenge that renders conventional IDMA inefficient for the unsourced MAC formulation in this paper. It is known that for the single user channel, generalized LDPC codes with Hadamard codes as check nodes (GHLDPC codes) exhibit close-to-capacity performance at very low rates~\cite{GLDPC}. Motivated by this result, we attempted to design rate-1/300 GHLDPC protograph codes for a multi-user channel using differential evolution; however, the optimization procedure did not iterate beyond initial population and the density evolution thresholds were poor. A better rate-1/300 code for IDMA was obtained by repeating each coded bit of a rate-1/4 LDPC code $75$ times. The minimum $E_b/N_0$ required to achieve a probability error of $0.05$ for this code is plotted in Fig.~\ref{fig:IDMA}.
 It can be seen that there is a significant gap between FBL bound and the performance of conventional IDMA, and that conventional IDMA scales very poorly with the number of users.  Our proposed scheme circumvents this code design bottleneck by sparsifying the transmissions and controlling the interference.
 It provides significant performance improvement at low complexity for a large number of users.  
 It is an interesting open problem to determine if there are other codes of rate-1/300 codes that could work well with conventional IDMA and without sparse repetition, even for a large number of users.

\section{Conclusion}
We proposed a CS and IDMA based scheme for the unsourced, uncoordinated Gaussian multiple access channel.
The difficulty of designing low-rate LDPC codes for IDMA is circumvented by introducing a sparse version of IDMA. 
We developed the density evolution equations for sparse IDMA with Gaussian approximation and designed protograph-based LDPC codes for sparse IDMA. When decoded with a message passing algorithm, the proposed coding scheme performs comparably to the other existing schemes in the large number of users regime.


\bibliographystyle{ieeetr}
\bibliography{IEEEabrv,MACcollision}
\end{document}